\makeatletter
\def\@oddfoot{\hbox{}\hfil \scriptsize \thepage}
\def\@evenfoot{\hbox{}\hfil \scriptsize \thepage}
\makeatother

\documentclass[conference]{IEEEtran}

\usepackage {cite}
\usepackage[]{algorithm2e}
\usepackage{graphicx}% http://ctan.org/pkg/graphicx
\usepackage{amsmath}
\usepackage{caption}
\usepackage{subcaption}
\usepackage{eso-pic}

\IEEEoverridecommandlockouts
\newcommand{\MYfooter}{\smash{
		\hfil\parbox[t][\height][t]{\textwidth}{\centering
			\thepage}\hfil\hbox{}}}

\makeatletter

\def\ps@IEEEtitlepagestyle{%
	\def\@oddhead{\parbox[t][\height][t]{\textwidth}{%\centering
			2016 8th International Symposium on Telecommunications (IST'2016)
		}\hfil\hbox{}}%
	\def\@evenhead{\scriptsize\thepage \hfil \leftmark\mbox{}}%
	\def\@oddfoot{ 978-1-5090-3435-2/16/\$31.00~\copyright~2016 IEEE \hfil 
		\leftmark\mbox{}}%
	\def\@evenfoot{\MYfooter}}

\makeatother
\begin{document}
\title{A Multi-Dimensional Fairness Combinatorial Double-Sided Auction Model in Cloud Environment}

\author{\IEEEauthorblockN{Reihaneh Hassanzadeh}
\IEEEauthorblockA{Department of Computer Engineering\\
	Sharif University of Technology
	\\Tehran, Iran\\
Email: rhassanzadeh@ce.sharif.edu}
\and
\IEEEauthorblockN{Ali Movaghar}
\IEEEauthorblockA{Department of Computer Engineering\\
	Sharif University of Technology
	\\Tehran, Iran\\
	Email: movaghar@sharif.edu}

\and
\IEEEauthorblockN{Hamid Reza Hassanzadeh}
\IEEEauthorblockA{Department of Comp. Science \& Eng.\\
	Georgia Institute of Technology\\
	USA, Georgia 30332\\
	Email: hassanzadeh@gatech.edu}
}

\maketitle

\begin{abstract}
In cloud investment markets, consumers are looking for the lowest cost and a desirable fairness while providers are looking for strategies to achieve the highest possible profit and return. Most existing models for auction-based resource allocation in cloud environments only consider the overall profit increase and ignore the profit of each participant individually or the difference between the rich and the poor participants. This paper proposes a multi-dimensional fairness combinatorial double auction (MDFCDA) model which strikes a balance between the revenue and the fairness among participants. We solve a winner determination problem (WDP) through integer programming which incorporates the fairness attribute based on the history of participants which is stored in a repository. Our evaluation results show that the proposed model increases the willingness of participants to take part in the next auction rounds. Moreover the average percentage of resource utilization is increased.
\end{abstract}

\IEEEpeerreviewmaketitle

\section{Introduction}
Cloud computing is an emerging technology in distributed environments where hardware and software computing resources are virtualized and delivered in the form of services to clients. Cloud computing provides a dynamic computing infrastructure, optimal resources allocation, rapid elastic access to computing architectures, reusable resources, flexibility, on-demand services, pay-per-use model, lower costs, and etc. \cite{pap1,pap2,pap3}. Therefore, due to many advantages of cloud computing, numerous users have been attracted to the use of cloud-based resources and services. The increasing number of users have made the management and allocation of cloud-based resources a challenging task. Resource allocation plays an important role in efficiency of the overall system and the level of consumer satisfaction of the deployed system. Therefore, it must be in such a way that in addition to delivering profit to the providers, consumer satisfaction will also be considered \cite{pap4}. Different approaches and models try to solve the resource allocation problem efficiently. Economics-based approaches such as market-based and auction-based resource allocation mechanisms have been two recent popular techniques to solve the aforementioned problem. In market-based approach, providers determine the resources price based on the number of resources that consumers are asking. While in the auction-based approach both providers and consumers have an impact on the agreed final prices \cite{pap5}.

\begin{figure}[t!]
	\centerline{
		\includegraphics[scale=0.6,trim=8cm 18.5cm 9cm 3cm]{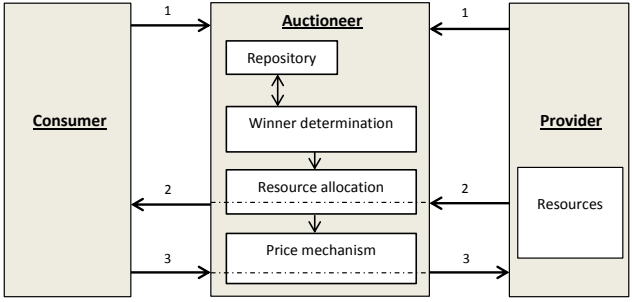}
	}
	\caption{A schematic view of an auction components and their interactions. In step 1 each consumer and provider send their bid to the auctioneer. After determination of the winners, in step 2 the auctioneer takes the resources from providers and gives them to the related consumers. In step 3 the transaction prices are specified and the consumers pay the payment to the providers.}
	\label{fig1}
\end{figure}

Auction is an efficient economic system based on bidding where buyer/vendor compete with each other to purchase/sell commodities \cite{pap3,pap6}. In an auction the vendors goal is to gain the most profit in the long run while the buyers expect their tasks to be run at the lowest cost with respect to the expected level of service. The auction system in the cloud environment is comprised of three major components, namely, the cloud provider (CP), the cloud consumer (CC) and the cloud auctioneer (CA). In Fig. \ref{fig1} the responsibility and the relationship between these three components is briefly specified. The auctioneer is responsible for allocating resources and determining the final price that is payable for each participant.

The bidder drop issue is one of the important concerns in auctions that should be dealt with. This problem occurs when participants with powerful and high quality bids have a better chance of winning the auction. As a result, weaker participants will be gradually pushed aside and the strong ones dominate the auction \cite{pap6}. Hence, adding a fairness term to the problem formulation when choosing the winners can help alleviate this issue by enforcing the equality conditions among participants. Even though the impact of such a policy may not be tangible in the short run, it tends to increase the willingness of participants to take part in the auctions in the long run \cite{pap9}.

Auctions in general can be placed in one of the major categories of one-sided or double-sided. In a one-sided auction, multiple buyers compete to purchase commodities of from vendor or multiple vendors compete with each other to sell their commodities to a single buyer. In contrast, in a double-sided auction, multiple buyers compete with multiple vendors to purchase the commodities. A double-sided auction has been shown to be more efficient than holding  multiple one-sided auctions \cite{pap7}. Moreover, an auction system can be either single-item or combinatorial. In a single item auction each bid is attributed to only one resource whereas in the combinatorial one, participants can request combinations of resources in their bids to run their tasks on. This type of auction is more efficient than old style single item auction \cite{pap7,pap8}.

In the rest of the paper, first a brief survey of the related works in multi-attribute auction domain is given focusing on approaches that integrate the fairness attribute into their model. In section 2 we introduce our proposed model in a combinatorial double auction which includes a fairness factor. In section 3 we evaluate our model and compare its performance with a similar auction model that disregards the individual fairness. Finally, in section 4 we conclude the paper. 

\section{Related Works}
There has been many studies that have tried to maximize the overall profits in some formulated auction resource allocation problem in a cloud environment as their main objective. Even though in some of these works, in addition to the price, factors such as delivery time, quality of service and even penalty policies are taken into account, but few of them have considered the fairness attribute.

In \cite{pap9} a fair multi-attribute combinatorial double auction model is introduced in order to increase the consumer satisfaction. Fairness and providers' reputation, are two attributes that are used by the authors of the article. In this model due to the policies that are considered, providers who give incorrect information will be punished and their reputation will be decreased in the next round of auction. Hence, this model seeks increasing user satisfaction and truthfulness of providers. In \cite{pap6} a multi-dimensional fairness mechanism is employed in an auction-based environment. In order to overcome the bidder drop problem, the author introduced three approaches of calculating of the fairness, namely, the quantitative, the qualitative, and the stochastic approaches. In their work a consumer which is in a miserable situation and as a result may leave the auction, is given a higher priority to reduce the possibility of him leaving the auction. 

On the other hand, the limited available resources as well as the high demand from users result in loss of fairness in the allocation of network resources. To address this issue, Wu et al. \cite{pap12} proposed an efficient relay resource allocation that considered fairness competition between users. Their method tried to add fairness by using the incomplete private information obtained from other nodes. In another study \cite{pap13}, authors designed an auction mechanism to allocate the spectrum resources with a trade-off between maximization of revenue and fairness. In their algorithm, bidders ranks will be changed based on the critical condition in order to decrease starvation. Finally, Lee et al. \cite{pap14} used a service oriented auction mechanism to deal with the bidder drop problem. In their proposed work, each resource has a reservation price and will be sold to the customer who has suggested a higher price. The resources that remain in the end will be shared among the weakest customers. 

\section{Auction Model and Problem Formulation}
In this section, we first describe our model and its fairness mechanism. Then the winner determination problem is formulated as an integer programming (IP) problem.
\subsection{Multi-Dimensional Fairness Combinatorial Double Auction Model (MDFCDA)}
In this article we propose a new combinatorial double auction in cloud environment that considers multi-dimensional fairness, henceforth, a multi-dimensional fairness combinatorial double auction model (MDFCDA). In our model we used multi-dimensional fairness inspired by work report in \cite{pap10} and formulate it in the form of an integer programming framework to solve the WDP and resource allocation problem.

In the proposed model, multiple providers, consumers and different resources (in terms of capacity and the number of CPU, memory size and bandwidth) are considered. Auction takes place in several rounds; in each round the providers and consumers individually deliver their bids to the auctioneer in order to sell and buy the resources. The providers offer the available resources in their bids and the consumers request for the bundle of resources in their bids to run their applications (tasks). The main task of the auctioneer after receiving and collecting the bids from different participants is first, to extend them and then to select the winners and allocate resources to them and determine the transaction price.

There is a repository for storing historical information of participants. In this repository, information such as the number of wins, the number of losses, the overall quality of the participant bids and other factors related to the previous auction round is stored. Hence, the auctioneer extends the bids and makes decision on determining the winners based on the information available in this repository. Moreover, by our assumption, if a resource is assigned to a consumer it will not be released until the end of the auction round and the execution time of the user tasks on the assigned resources ends by then. 

The fairness attribute helps reducing the number of bidders that drop and as a result the number of participants who could not compete with more powerful participants and leave the auction will be reduced. A. Pla \cite{pap6}, has introduced a quantitative, a qualitative and a stochastic approach for enforcing the fairness. In a quantitative approach, fairness is calculated based on the number of victories and failures of participants in the previous rounds of auction. This approach encourages the weak participants to participate in the upcoming auction rounds by increasing the chances of their winning. But the main issue with this approach is that each participant can easily increases his chance of winning with a dummy bid. A dummy bid is a bid which has the lowest possible price for the resources to increase the odds of winning the next rounds after losing several rounds. To solve this problem and also to affect the bids attributes to increase or decrease the chance of winning, the qualitative approach is proposed.  In contrast, in a quantitative approach the chance of winning for each participant depends on the bid attributes (e.g. the suggested price, QoS, etc). Finally, in a stochastic approach, measurements that are performed based on the previous two approaches are randomly applied.

In our proposed approach, fairness attribute is guaranteed by a fairness factor that takes on a real value. The higher the value, the more the chance of winning will be. This defined fairness factor fuses all the previously mentioned perspectives (i.e. qualitative, quantitative and stochastic approaches) into one single measure. In cases where two participants have similar conditions in terms of the number of wins and losses, the qualitative approach favors the one with a higher bid value. Therefor the fairness factor is calculated by the qualitative and quantitative approaches and then it is tuned according to the stochastic approach.

The following policies are considered in computing the fairness factor:
\begin{itemize}
	\item On the one hand, the participants who have more losses are given more chance of winning and on the other hand, participants who have more wins are given less chance of winning.
	\item Among the losers of previous auction round, the participant who has the bid with higher quality (such as having higher bid price, lesser service time, etc.) are given more chance of winning. Also among the winners of previous auction round, the participant who has the bid with lower quality are given less chance of winning.
	\item The more the participant be in on the verge of a drop (i.e. has lost a higher number of consecutive rounds) the higher the value that the fairness factor will take on. 
\end{itemize}

Algorithm 1 shows the procedure to calculate the fairness factor for consumers. In this algorithm $CL_n$ is the last number of consecutive losses for the consumer $n$ (assuming that after each victory, $CL_n$ takes on value 0). $ml$ is the maximum consecutive auction rounds that consumers can loose and if $CL_n>ml$ the bidder drop problem occurs. Based on this algorithm if the consumer has participated in the current round two possible scenarios can happen:
\begin{itemize}
	\item 	If the consumer $n$ lost in the previous auction round, fairness factor will be applied with probability of $Prob\_W$. The higher $CL_n$ becomes the more $Prob\_W(CL_n)$ will be. Hence a consumer who is in a more critical condition to drop the auction has a higher chance of winning. Fairness factor is hence calculated according to the following equation:
\begin{equation}
\label{eq3}
\begin{split}
Fun\_W (\#losses_n,eval\_fun_n,CL_n)=\\ 
(CL_n+1)\times ((\alpha_1\times \#looses_n)+(\alpha_2\times eval\_fun_n))
\end{split}
\end{equation}
$\#looses_n$ is the total number of looses in the previous auction rounds and also $eval\_fun_n$ is evaluation function to determine the quality of previous auction rounds' bids of consumer $n$ based on the historical information (e.g. the number of requested resources, suggested prices, etc.) and will be calculated through the equations that is introduced in \cite{pap10}. Where $\alpha_1$ and $\alpha_2$ coefficients are factors to strike a balance between the aforementioned two terms.
\item 	If the consumer $n$ won in the previous auction round, fairness factor will be applied with the probability of $Prob\_L$. The smaller $CL_n$ becomes the higher $Prob\_L(CL_n)$ and hence, the smaller chance of winning for a consumer. For this case, the fairness factor is calculated according to the following equation:
\begin{equation}
\label{eq4}
\begin{split}
Fun\_L(\#wins_n,eval\_fun_n,CL_n)=\\
-1\times\frac{1}{CL_n+1}\times ((\beta_1\times \#wins)+\frac{\beta_2}{eval\_fun_n})
\end{split}
\end{equation}

\end{itemize}
where $\#wins$ is the total number of wins in the previous auction rounds and $\beta_1$ and $\beta_2$ play similar roles to $\alpha_1$ and $\alpha_2$.

\begin{algorithm}
 \KwResult{Fairness factor}
  \For {consumer$_n \in $ participants\_in\_current\_auction\_round}{
	  \eIf {consumer$_n$\_failed\_in\_previous\_auction\_round }{
		  \If {$Random(0,1) < Prob\_W(CL_n)$}{
			 FairnessFactor$_n$= $Fun\_W(\#looses_n,eval\_func_n,CL_n)$
		  }
	  }{
      \If {$Random(0,1) < Prob\_L(CL_n)$}{
      	FairnessFactor$_n$= $Fun\_L(\#wins_n,eval\_func_n,CL_n)$
      }
    }
 }
 \caption{Calculation of the fairness factor algorithm}
\label{alg1}
\end{algorithm}

\section{Problem Formulation}
The proposed model consists of three main phases of 1) bidding policies, 2)winning determination problem (WDP) and 3) pricing mechanism. In the bidding policies phase, there is a way each participant should bid according to the predefined rules. In phase of WDP phase, winners are determined with the aim of maximizing the total utility considering fairness factor using integer programming. After determining the winners, in the pricing phase trade prices that consumers should pay to providers are calculated.

\subsection{Bidding Policies}
In the proposed model $N$ consumers, $M$ providers and $L$ resource types are considered. Each of the consumers and the providers request and offer different types of resources for buying and selling, independently from each other. The consumer $n (1\le n \le N)$ requests the following two-tuple bid:

\begin{itemize}
	\item $Bid_n^{'}  = (Price_n^{'},Quantity_n^{'})$, where $Price_n^{'}=(P_{n_1}^{'},P_{n_2}^{'},\dots,P_{n_L}^{'})$ is the suggested prices for the requested resources by the consumer $n$, $P_{n_l}^{'}$ is the suggested price per unit of resource type $l$ $(1\le l \le L)$ by the consumer $n$, $Quantity_n^{'}=(Q_{n_1}^{'},Q_{n_2}^{'},\dots,Q_{n_L}^{'})$ is the required resources bundle by the consumer $n$ and $Q_{n_l}^{'}$ indicates the required resource type $l$ by the consumer $n$.
\end{itemize}
Also The provider $m$, $1\le m \le M$ offers the following two-tuple bid:
\begin{itemize}
	\item 	$Bid_m  = (Price_m, Quantity_m )$, where $Price_m=(P_{m_1},P_{m_2}\dots,P_{m_L} )$ is the suggested prices for the offered resources by the provider $m$, $P_{m_l}$ indicates the suggested price per unit of resource type $l$ by the provider $m$, $Quantity_n=(Q_{m_1},Q_{m_2},\dots,Q_{m_L} )$ is the offered resources bundle by the provider $m$ and $Q_{m_l}$ indicates the offered resource type $l$ by the provider $m$.
\end{itemize}
Next, the auctioneer collects the initial bids of the consumers and then according to the stored historical information in the repository extends the bids by fairness attribute. We defined the consumer extended bid as $BidExt_n^{'}=(Bid_n^{'},FairnessFactor_n )$ where $FairnessFactor_n$ is the fairness factor for the consumer $n$ calculated by Algorithm 1.

\subsection{Winning Determination Problem (WDP)}
Winner determination problem (WDP) plays an important role in an auction and its performance will change according to which algorithm is used \cite{pap3}. The purpose of WDP is to determine the participants who have the best bids as the winners. The best bids are selected based on criteria such as the suggested price for each unit of the resources. In our proposed method, we formulate WDP in integer programming (IP) framework to maximize the total utility of the participants. Also in order to guarantee fairness in the auction we apply the fairness factor based on the criteria mentioned previously in a random fashion. Applying the fairness factor besides maximization of the utility, decreases the chance of winning of a participant who has dominated the auction and by contrast, increases the winning odds for the weaker participants. Accordingly, we define the objective function as follows: $Obj = \max$ \{Total Utility + Total Satisfaction\}.

Each provider and consumer has a utility function which represents the difference between the suggested price and the trade price and is calculated according to \eqref{eq5} for the provider $m$ and the consumer $n$.
\begin{equation}
\begin{split}
Utility_m = trade\;price - suggested\;price\\
Utility_n = suggested\; price - trade\; price
\end{split}
\label{eq5}
\end{equation}
The Total Utility of participants is the summation of all consumers and providers utilities and is obtained through the following formula:
\begin{equation}
\begin{split}
Total\; Utility=Consumers\; Utilities+Providers\;Utilities=&\\
(\sum_{n,l} x_n  P_{n_l}^{'} Q_{n_l}^{'} - \sum_{n,l,m}y_{nlm} TradePrice_{nlm} )+& \\
(\sum_{n,l,m} y_{nlm} (TradePrice_{nlm}- P_{ml}))&
\end{split}
\label{eq6}
\end{equation}
\begin{equation}
v_n = \sum_l P_{nl}'Q'_{nl}
\label{eq7}
\end{equation}

In the above equations, $TradePrice_{nlm}$ is the transaction price between the consumer $n$ and provider $m$ per unit of resource type $l$. Also $v_n$ is the total budget that the consumer $n$ can pay for all the required resources. Accordingly, we have,
\begin{equation*}
Total\; Utility=\sum_n x_n.v_n -\sum_{m,n,l} y_{nlm}.P_{ml} 〗
\end{equation*}
and the Total Satisfaction of consumers is computed as:
\begin{equation*}
Total\; Satisfaction = \left(\sum_n x_n FairnessFactor_n \right)
\end{equation*}
where the constraints are tied together according to the following integer programming formulation:	
\begin{eqnarray*}
x_n\in\{0,1\}  & \forall n \in 1\dots N\\
y_{nlm} \in \{0,1,\dots,Q_{ml} \}   &  \forall m \in 1\dots M ,\\
 &\forall n \in 1\dots N, \forall l \in 1\dots L\\
x_n \ge \left( \sum_{l,m} y_{nlm} \right)/\left(\sum_l Q_{nl}^{'} \right)  & \forall n \in 1\dots N\\
x_n \le \sum _ {l,m} y_{nlm} & \forall n \in 1\dots N\\
\sum_n y_{nlm}  \le Q_{ml}   & \forall l \in 1\dots L,\\
& \forall m \in 1\dots M\\
\sum_m y_{nlm} = x_n.Q_{nl}^{'}   & \forall  n \in 1\dots N, \\
&\forall l \in 1\dots L\\
y_{nlm}\left(P_{nL}^{'} - P_{mL}\right)\ge 0 & \forall  n \in 1\dots N ,\forall l \in 1\dots L,\\
& \forall m \in 1\dots M\\
\end{eqnarray*}
In the above relations, $x_n=1$ indicates that consumer $n$ has won the auction round and gets all the resources he requested in his bid and lost otherwise ($x_n=0$), $y_{nlm}$ is the number of resource type $l$ sold by the provider $m$ to consumer $n$ and $v_n$ is the suggested price for all requested resources by the consumer $n$.

\subsection {Pricing Mechanism}
After determining the winners of the auction, each consumer must pay an amount to use resources of the related provider. This amount should be fair for both the customer and the provider to bring about some level of satisfaction among both parties. In the calculation of trade price, the average suggested prices approach is considered for each type of resources. It is assumed that consumer $n$ received the resources of type $l$ from provider $m$ and therefore the final payable amount for this resource is given by:
\begin{equation}
TradePrice_{nlm}=\frac{(P_{nl}^{'}+P_{ml})}{2}
\label{eq17}
\end{equation}
Moreover, the total payment that consumer $n$ (provider $m$) must pay (receives) is computed as bellow:
\begin{equation}
TradePrice_{n}=\sum_{m=1}^{M}\sum_{l=1}^LTotal\_q_{nlm}TradePrice_{nlm}
\label{eq18}
\end{equation}
\begin{equation}
TradePrice_{m}=\sum_{n=1}^{N}\sum_{l=1}^LTotal\_q_{nlm}TradePrice_{nlm}
\label{eq19}
\end{equation}
where in the above equation $Total\_q_{nlm}$ is the number of resources of type $l$ that consumer $n$ takes from provider $m$.

\section{Performance Evaluation }
In this section, we simulate our proposed model to evaluate and compare it with a model which do not incorporate the fairness factor into its model, in terms of utility, the average percentage of resource utilization, the average percentage of winnings for consumers, the number of bidder drops, and the average round that bidder drop occurs. We use the Java language to implement the model and ILOG CPLEX v12.6 library to solve the corresponding WDP.

\subsection{Experimental setup}
We simulate an auction for 100 rounds and we repeated it for 10 times with different random seeds. We considered 4 different types of resources, 300 consumers and 5 providers. Each provider offers a random number (according to a uniform distribution, $U(30,100)$) of resource for each resource category, and each consumer requests from each type of resource 1-3 times (according to a uniform distribution, $U(1,3)$) in range of [1, 3]. Also the suggested price of each provider is distributed randomly within the range $[50, 200]$ and the suggested price for each consumer is selected, randomly ($U[100,250]$), according to the suggested prices from the previous round. For each consumer, the bidder drop problem occurs when he loosed 7 consecutive auction rounds in a row. Table \ref{tab1} demonstrates the selected parameters for the described scenario. 
\begin{table} 
	\centering
	\caption{Algorithm parametrization}
	\label{my-label}
	\begin{tabular}{||c|c||}
		\hline
		Parameter name     & Parameter value \\\hline\hline
		$\alpha_1$         & 9              \\\hline
		$\alpha_2$         & 7               \\\hline
		$\beta_1$          & 4               \\\hline
		$\beta_2$          & 28              \\\hline
		$ml$               & 6               \\\hline
		$N$                & 300               \\\hline
		$M$                & 5               \\\hline
		$L$                & 4               \\\hline
		\# rounds          & 100             \\\hline
		\# simulation runs & 10             \\\hline
	\end{tabular}
	\label{tab1}
\end{table}

\subsection{Experimental Result}
\begin{figure}[t!]
	\centering
	\includegraphics[scale=0.2,trim=0.5cm 3.5cm 0.5cm 4cm]{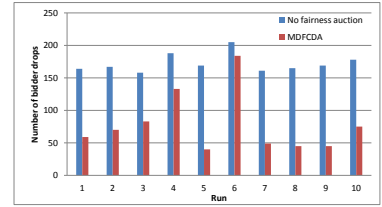}
	\caption{The number of bidder drops across different runs }
	\label{fig2}
\end{figure}

\begin{figure}
	\centering
	\includegraphics[scale=0.2,trim=0.5cm 3.5cm 0.5cm 4cm]{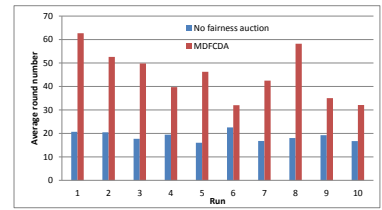}
	\caption{The average round number that bidder drop occurs across different runs.}
	\label{fig3}
\end{figure}

\begin{figure}[t]
	\centering
	\includegraphics[scale=0.2,trim=0.5cm 3.5cm 0.5cm 4cm]{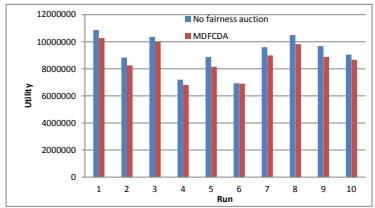}
	\caption{The total utility across different runs.}
	\label{fig4}
\end{figure}

\begin{figure}[t]
	\centering
	\includegraphics[scale=0.2,trim=0.5cm 3.5cm 0.5cm 4cm]{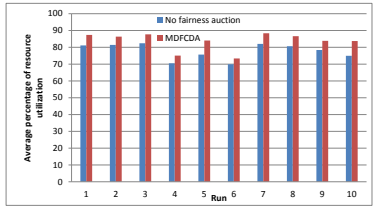}
	\caption{The average percentage of resource utilization across different runs.}
	\label{fig5}
\end{figure}

\begin{figure}
	\centering
	\includegraphics[scale=0.23,trim=0.5cm 4.5cm 0.5cm 3.5cm]{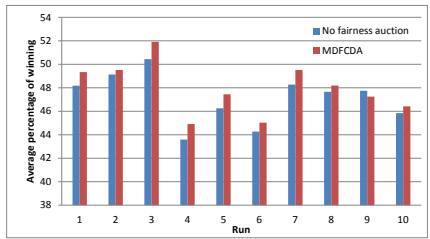}
	\caption{The average percentage of winning across different runs.}
	\label{fig6}
\end{figure}

\begin{figure}[t!]
	\centering
	    \includegraphics[scale=0.23,trim=0.5cm 4.5cm 0.5cm 3.5cm]{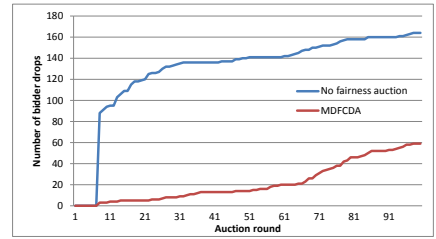}
	    \caption{The number of bidder drops across different rounds of the first run.}
   	    \label{fig7}
\end{figure}

\begin{figure}
		\centering
	    \includegraphics[scale=0.23,trim=0.5cm 4.5cm 0.5cm 3.5cm]{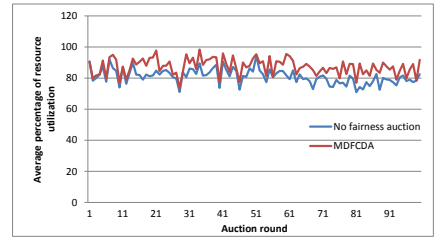}
	    	    \caption{The average percentage of resource utilization across different rounds of the first run.}
	    \label{fig8}
 \end{figure}

Figure \ref{fig2} and \ref{fig3} make comparisons between the MDFCDA and the no fairness models in terms of the number of bidder drops and the average round number that bidder drop occurs. As depicted in these figures, the number of drops in MDFCDA is consistently smaller than the baseline model and the bidder drop always occurs in the later rounds. Figure \ref{fig4} depicts the total utility (i.e. summation of providers and consumers utilities) in different runs of auction. According to the figure, in terms of the utilities the baseline model slightly outperforms the MDFCDA which is a trade-off due to the higher fairness in the MDFCDA model. Figure \ref{fig5} and \ref{fig6} compare the average percentage of resource utilization and the average percentage of consumer wins between the two models. As clearly implied by the figures, the proposed model outperforms the baseline model. Figure \ref{fig7} and \ref{fig8} illustrate cumulative sum of the number of bidder drops and the average percent of resource utilization, for a single run respectively. Overall, from the simulation results we can conclude that our model improves the average percentage of resource utilization and the average percentage of consumer wins and also decreases the number of bidder drop in different auction rounds.


\section{Conclusion and Future work}
Fair resource allocation has been always a challenging task in auction systems, yet few papers have addressed this important criterion in auction systems. Here we made an effort to integrate the fairness as a factor by leveraging the power of integer programming and efficient tools to solve the optimum solutions for this NP-complete problem. In so doing, we developed a combinatorial double-sided auction model with fairness factor to maximize the overall profit while at the same time maintaining some level of fairness. Our results showed that our proposed model promisingly outperforms a similar baseline model where the fairness factor is not considered. The MDFCDA model can efficiently deal with the bidder drop problem and yield better resource utilization. Moreover, It results in a higher average number of winners for most cases.

\bibliographystyle{abbrv}
\bibliography{Paper}

\begin{thebibliography}{10}

\bibitem{pap9}
G.~Baranwal and D.~P. Vidyarthi.
\newblock A fair multi-attribute combinatorial double auction model for
  resource allocation in cloud computing.
\newblock {\em Journal of Systems and Software}, 108:60--76, 2015.

\bibitem{pap4}
A.~Beloglazov, J.~Abawajy, and R.~Buyya.
\newblock Energy-aware resource allocation heuristics for efficient management
  of data centers for cloud computing.
\newblock {\em Future generation computer systems}, 28(5):755--768, 2012.

\bibitem{pap5}
I.~Fujiwara, K.~Aida, and I.~Ono.
\newblock Applying double-sided combinational auctions to resource allocation
  in cloud computing.
\newblock In {\em Applications and the Internet (SAINT), 2010 10th IEEE/IPSJ
  International Symposium on}, pages 7--14. IEEE, 2010.

\bibitem{pap14}
J.-S. Lee and B.~K. Szymanski.
\newblock A novel auction mechanism for selling time-sensitive e-services.
\newblock In {\em E-Commerce Technology, 2005. CEC 2005. Seventh IEEE
  International Conference on}, pages 75--82. IEEE, 2005.

\bibitem{pap2}
P.~Mell and T.~Grance.
\newblock The nist definition of cloud computing.
\newblock 2011.

\bibitem{pap10}
A.~Pla, B.~Lopez, and J.~Murillo.
\newblock Multi criteria operators for multi-attribute auctions.
\newblock In {\em Modeling Decisions for Artificial Intelligence}, pages
  318--328. Springer, 2012.

\bibitem{pap6}
A.~Pla, B.~L{\'o}pez, and J.~Murillo.
\newblock Multi-dimensional fairness for auction-based resource allocation.
\newblock {\em Knowledge-Based Systems}, 73:134--148, 2015.

\bibitem{pap1}
X.~Wang, X.~Wang, C.-L. Wang, K.~Li, and M.~Huang.
\newblock Resource allocation in cloud environment: a model based on double
  multi-attribute auction mechanism.
\newblock In {\em Cloud Computing Technology and Science (CloudCom), 2014 IEEE
  6th International Conference on}, pages 599--604. IEEE, 2014.

\bibitem{pap13}
C.~Wu, S.~Zhong, and G.~Chen.
\newblock A strategy-proof spectrum auction for balancing revenue and fairness.
\newblock In {\em Consumer Communications and Networking Conference (CCNC),
  2014 IEEE 11th}, pages 827--832. IEEE, 2014.

\bibitem{pap12}
D.~Wu, Y.~Cai, and M.~Guizani.
\newblock Auction-based relay power allocation: Pareto optimality, fairness,
  and convergence.
\newblock {\em Communications, IEEE Transactions on}, 62(7):2249--2259, 2014.

\bibitem{pap7}
M.~Xia, J.~Stallaert, and A.~B. Whinston.
\newblock Solving the combinatorial double auction problem.
\newblock {\em European Journal of Operational Research}, 164(1):239--251,
  2005.

\bibitem{pap3}
K.~Xu, Y.~Zhang, X.~Shi, H.~Wang, Y.~Wang, and M.~Shen.
\newblock Online combinatorial double auction for mobile cloud computing
  markets.
\newblock In {\em Performance Computing and Communications Conference (IPCCC),
  2014 IEEE International}, pages 1--8. IEEE, 2014.

\bibitem{pap8}
S.~Zaman and D.~Grosu.
\newblock Combinatorial auction-based allocation of virtual machine instances
  in clouds.
\newblock {\em Journal of Parallel and Distributed Computing}, 73(4):495--508,
  2013.

\end{thebibliography}

\end{document}